\tolerance   10000
\magnification = 1200
\baselineskip=1.65\normalbaselineskip
\font\ti=cmti10  scaled 800
\font\tibig=cmti10 scaled 1000
\font\bigrmtwenty = cmb10 scaled 1500 
\font\bigrmsixteen = cmb10 scaled 1000
\font\smallrm = cmr10 scaled 800

\font\name = cmr10 scaled 1200
\input epsf
\let\origeqno=\eqno
\def\eqno(#1){\origeqno (\rm #1)}

\noindent
{\centerline {\bigrmtwenty  PARTIAL AVERAGING NEAR A RESONANCE IN    }}
\hfill
\noindent 
{\centerline {\bigrmtwenty   PLANETARY DYNAMICS }}
\vskip   15pt
\noindent
{\centerline  {\bf{\name NADER HAGHIGHIPOUR $^{^\ast}$}}}
\hfill
\noindent
{\centerline{\ti {Department of Physics and Astronomy, 
             University of Missouri-Columbia, Columbia ,}}}
\hfill
\noindent
{\centerline {\ti {Missouri$\>$ 65211 , U.S.A.}}}
\vskip  10pt
\noindent
{\bigrmsixteen  Abstract.}
Following the general numerical analysis of Melita and Woolfson (1996),
I showed in a recent paper that a restricted, planar, circular planetary system 
consisting of Sun, Jupiter and Saturn would be captured in a near (2:1) resonance 
when one would allow for frictional dissipation due to interplanetary medium 
(Haghighipour, 1998). In order to analytically explain this resonance phenomenon,
the method of partial averaging near a resonance was utilized and the dynamics 
of the first-order partially averaged system at resonance was studied. Although 
in this manner, the finding that resonance lock 
occurs for all initial relative positions of Jupiter and Saturn was confirmed, 
the first-order partially averaged system at resonance did not provide a complete 
picture of the evolutionary dynamics of the system and the similarity between
the dynamical behavior of the averaged system and the main planetary system held 
only for short time intervals. To overcome these limitations, the method of partial 
averaging near a resonance is extended to the second order of perturbation in this
paper and a complete picture of dynamical behavior of the system at resonance
is presented. I show in this study that the dynamics of the second-order partially 
averaged system at resonance resembles the dynamical evolution of
the main system during the resonance lock in general, and I present analytical 
explanations for the evolution of the orbital elements of the main system while 
captured in resonance.
\vskip  10pt
\noindent
{\bigrmsixteen  Key words:} planetary dynamics , resonance capture , 
                                                    averaging .
\vskip  15pt
\noindent
$^\ast$E-mail:nader@lula.physics.missouri.edu 
\vfill
\eject
\noindent
{\centerline {\bigrmsixteen {1. $\>$ Introduction}}}
\noindent
\vskip  5pt

Inspired by the results of extensive numerical investigations by Melita
and Woolfson (1996) and in the framework of a restricted planar circular
three-body system, I have recently shown that a planetary system
consisting of a star and two planets, subject to dynamical friction
due to a freely rotating uniform interplanetary medium, will be captured in
resonance when the inner planet is more massive (Haghighipour, 1998,
hereafter paper I). As an example of such a planetary system, the three-body
system of Sun-Jupiter-Saturn was numerically integrated and it was shown
that the orbital period of Saturn would be captured in a near (2:1)
commensurability with that of Jupiter (Figure 1). In this system, the motion of 
Sun was neglected and Jupiter was assumed to be moving on a circular
orbit with a known constant frequency. As a result of this resonance trapping, 
the eccentricity and semimajor axis of Saturn's osculating ellipse and 
also its orbital angular momentum and total energy became essentially constant 
(paper I).
\vskip  20pt
\hskip  70pt
\epsfbox{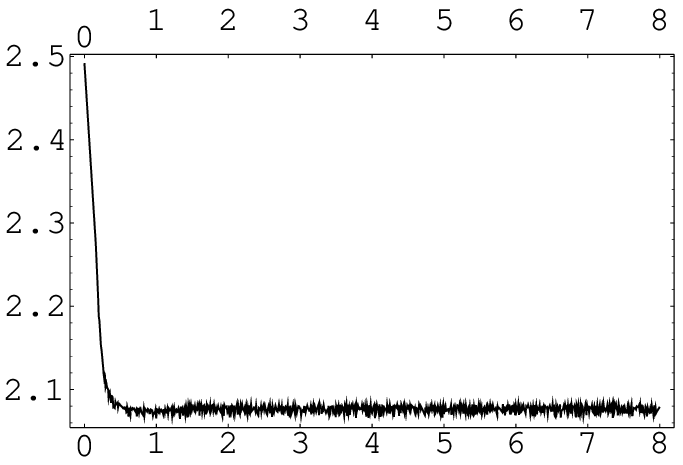}
\hfill
\vskip  20pt
\vbox{
\baselineskip=\normalbaselineskip
\smallrm
\noindent
{\ti Figure 1.} {\smallrm { Graph of the ratio of orbital period of Saturn
to that of Jupiter versus time. The system is captured in a near (2:1)
resonance. Integrations were performed on a timescale of ${10^4}({T_{_J}}/{2\pi})$ 
years where $T_{_J}$ is the orbital period of Jupiter. The system was integrated 
with Sun at the origin of an inertial plane-polar coordinate system, Jupiter 
initially on the x-axis and Saturn initially at 
$(a,e,\theta,{\hat v})=(1.8380462,0.0556,0,0)$ where $a$ and $e$ are the semimajor
and the eccentricity of Saturn's osculating ellipse, respectively, and $\hat v$ is 
its true anomaly.
The density of the interplanetary medium was taken to be equal to
16 times the mass of Jupiter uniformly spread in a spherical volume with 50 au 
radius and the masses of Sun, Jupiter and Saturn were taken to be constant and 
equal to their present values.}}}
\vskip  40pt

This paper is devoted to the analysis of the dynamical behavior of the planetary 
system described above while captured in resonance. A preliminary analysis of the 
dynamics of this system was presented in paper I using the principle of averaging ; 
in fact in paper I, the first-order 
partially averaged system at resonance was obtained and its dynamics was studied. 
It turned out that in the first order of approximation, the partially averaged 
system was Hamiltonian with a periodic potential function that guaranteed the 
occurrence of resonance lock for all initial relative positions of the two planets.
The dynamics of the first-order averaged system agreed with that of the main
three-body system only for short time intervals and it was mentioned that in 
order to obtain an agreement which would hold for longer times, one had to
extend the analysis to the study of the dynamics of the second-order partially 
averaged system at resonance. 

In this paper, a complete picture of the dynamical behavior of the three-body 
system of Sun-Jupiter-Saturn during resonance capture is presented. This is accomplished
by extending the calculations to the second-order partially averaged system at
resonance, where one can also explain the behavior of Saturn's orbital elements 
and angular momentum during the resonance lock. 

The model under investigation is presented in section 2. Section 3 contains a brief
discussion of the method of partial averaging near a resonance and in
section 4, the application of this method to the Sun-Jupiter-Saturn system is
discussed.  Section 5 concludes this study by reviewing and summarizing the results.
\vskip  20pt
\noindent
{\centerline {\bigrmsixteen {2. The Model}}} 
\vskip  5pt

The model under investigation is a restricted planar circular
three-body system consisting of Sun, Jupiter and Saturn. In this
model, the motion of Sun due to gravitational attraction of Jupiter
and Saturn is neglected and Jupiter is assumed to have a uniform circular
motion with a known constant period $T_{_J}$. The only 
source of dissipation in this system is the dynamical friction due to 
interplanetary medium. It has been shown by Dodd and McCrea (1952) and also by Binney 
and Tremaine (1987) that the effect of the dynamical friction on the orbital motion 
of a general spherical body $b$ with mass $m_{_b}$ around a general star with mass
$M$ appears as a deceleration with a magnitude given by
$$
R\,=\,{{2\pi{{\cal G}^2} {\rho_{_0}}}\over {V_{rel}^2}}\,{m_{_b}}\>
\ln\Bigl[1\,+\,{\bigl({{S\,{V_{rel}^2}}\over {{\cal G}\,{m_{_b}}}}
\bigr)^2}\Bigr]\>\>\>,
\eqno (1)
$$
\noindent
and a direction that is opposite to the velocity of body $b$ with respect to the medium.
In this equation, $\rho_{_0}$ is the uniform density of the interplanetary 
medium, $\cal G$ is Newton's constant and $S={r_{_b}}{({m_{_b}}/{2M})^{1/3}}$, 
where $r_{_b}$ is the radial distance of $b$ to the central star.
The relative velocity ${\vec V}_{rel}$ in
equation (1) is given by ${{\vec V}_{rel}}={\vec V}-{{\vec V}_m}$, where $\vec V$
is the velocity of the body and ${\vec V}_m$ is that of the interplanetary medium
at the position of the body. For the planetary system presented in this paper, it is 
assumed that the interplanetary medium is freely rotating around the central star
in the same sense as body $b$. Therefore, ${\vec V}_m$ is perpendicular to the
position vector of $b$ and has a magnitude equal to $(GM/{r_{_b}})^{1/2}$.

In an inertial coordinate system with its origin on Sun, the equation of
motion of Saturn in dimensionless form is given by(paper I)
$$
{{{d^2}{\vec r}}\over {d{t^2}}}\,=\,-\,{{\vec r}\over {r^3}}\,-\,
\varepsilon\>{{({\vec r}-{{\hat r}_{_J}})}\over {|{\vec r}-
{{\hat r}_{_J}}|^3}}\,-\,{\vec {\cal R}}\>\>\>.
\eqno  (2)
$$
\noindent
In this equation, $\varepsilon={m_{_J}}/M$, where $m_{_{_J}}$ is the mass of Jupiter,  
${\hat r}_{_J}$ is the unit vector along Jupiter's orbital radius and
${\vec {\cal R}}={\cal R}{{\vec {\cal V}}_{rel}}/{{\cal V}_{rel}}$ where
$$
{\cal R}\,=\,{A\over {{\cal V}_{rel}^2}}\,\ln(1\,+\,B{r^2}{{\cal V}_{rel}^4})\,,
\eqno (3)
$$
\noindent
is the dimensionless magnitude of Saturn's deceleration due to 
dynamical friction and
${\cal V}_{rel}$ is dimensionless magnitude of its
velocity with respect to interplanetary medium and is given by 
$$
{{\cal V}_{rel}^2}\,=\,{{\dot r}^2}\,+\,{r^2}{({\dot \theta}\,-\,
{\omega_m})^2}\,,
\eqno  (4)
$$
\noindent
where ${\omega_m}={r^{-3/2}}$, is the dimensionless angular 
frequency of the medium at distance $r$ from Sun. 
Denoting the mass of Saturn by ${m_{_S}}$, one finds that
$A=2\pi ({\rho_{_0}}{r_{_J}^3}/M)({m_{_S}}/M)$
and $B={2^{-2/3}}{({m_{_S}}/M)^{-4/3}}.$ For any physical system $A<<1$ and
$\delta\equiv AB<<1$. It is important to mention that in deriving equation
(2) a set of units has been chosen such that
${({T_{_J}}/2\pi)^2}={r_{_J}^3}/{\cal G}M$ and
all positions and time variables have been scaled by the orbital 
radius of Jupiter $(r_{_J})$ and its orbital period divided by $2\pi$, respectively 
(see paper I for details).

\vskip  10pt
\noindent
{\centerline {\bigrmsixteen {3. $\>$ Partially Averaged System at Resonance}}}
\vskip  10pt

In order to apply the method of partial averaging near a resonance to the problem
presented in this paper, it is useful to write equation (2) in terms of 
the Delaunay action-angle variables (Appendix A). Denoting the semimajor
axis and the eccentricity of Saturn's osculating ellipse by $a$ and $e$
respectively, the appropriate Delaunay variables for the relative motion (2)
are given by $L={a^{1/2}}, G=L{(1-{e^2})^{1/2}}, l={\hat u}-e\sin {\hat u}\,$ and 
$\,g=\theta-{\hat v}$, where ${\hat u}\,,{\hat v}$ and $l$ are the eccentric, the 
true and the mean anomalies of Saturn's orbit, 
respectively(Kovalevsky, 1967; Sternberg, 1969; Hagihara, 1972). In these relations,
$L$ and $G$ are action variables and $l$ and $g$ are their corresponding angular
variables. From Appendix A, it is evident that For the pure Kepler problem, i.e. 
in the absence of perturbations in 
equation (2), the solutions of the Delaunay equations are given by constant $L\,,G$
and $g$ and
$$
l\,=\,{\omega_{_K}}t\,+\,{l_{_0}}\>\>,
\eqno  (5)
$$
where ${\omega_{_K}}={L^{-3}}$ is the Keplerian frequency and $l_{_0}$ is a constant
of integration.

The problem under investigation here is a Kepler system with period 
${\cal T}=2\pi/{\omega_{_K}}$ that is perturbed by forces that involve a period
${\cal T}_p$. At resonance, it is necessary, but not sufficient that $\cal T$ and
${\cal T}_p$ become commensurate, i.e. relatively prime integers $n$ and $n'$
exist such that $n{{\cal T}_p}={n'}{\cal T}$. To explain the dynamical behavior
of the system near the resonance manifold in phase space, it is more convenient to
imagine a system with a single degree of freedom and action-angle variables
$({\cal I},{\Theta})$ such that
\vskip  6pt
$$
{\dot {\cal I}}\,=\,\epsilon \, {\cal P}({\cal I},\Theta,t,\epsilon)\,,
\eqno  (6)
$$
\noindent
and
$$
{\dot \Theta}\,=\,\Omega({\cal I})\,+\,
\epsilon\,{\cal N}({\cal I},\Theta,t,\epsilon)\,,
\eqno (7)
$$
\noindent
where $\cal P$ and $\cal N$ are explicitly time-dependent with period ${\cal T}_p$.

In the absence of perturbation $(\epsilon = 0)$, equations (6) and (7) represent a 
Hamiltonian system whose action variable $\cal I$ has a constant value 
${\cal I}_{_0}$. At resonance, this action variable fluctuates around its unperturbed 
value. Let us denote these fluctuations by
$$
{\cal I}\,-\,{{\cal I}_{_0}}\,=\,{\epsilon'}\,{\cal D}\,,
\eqno (8)
$$
\noindent
where $\epsilon'$ is a small parameter which has yet to be determined.
In a similar fashion, let us show deviations of $\Theta$ from its unperturbed value
$\Omega({{\cal I}_{_0}})t$ by
$$
\Theta\,-\,\Omega({{\cal I}_{_0}})\,t\,=\,\Phi\,.
\eqno (9)
$$
\noindent
Substituting for $\cal I$ and $\Theta$ from equations (8) and (9) in equations
(6) and (7), the dynamical equations of the system will be given by
\vskip  5pt
$$
{\dot {\cal D}}\,=\,{\epsilon \over {\epsilon'}}\,
{\cal P}({{\cal I}_{_0}},\Theta,t)\,+\,\epsilon\,{\cal D}
{{\partial {\cal P}} \over {\partial {\cal I}}}\,({{\cal I}_{_0}},\Theta,t)\,+\,
O(\epsilon {\epsilon'})\,,
\eqno (10)
$$
\noindent
and
$$
{\dot \Phi}\,=\,{\epsilon'}\,{\cal D}\,{{\partial \Omega}\over {\partial {\cal I}}}
({{\cal I}_{_0}})\,+\,\epsilon\, {\cal N}\,({{\cal I}_{_0}},\Theta,t)\,
+\,{1\over 2}\,{{\epsilon'}^2}\,{{\cal D}^2}\,
{{{\partial ^2} \Omega}\over {\partial {{\cal I}^2}}}({{\cal I}_{_0}})\,
+\,O(\epsilon {\epsilon'})\,+\,O({{\epsilon'}^3})\,.
\eqno  (11)
$$
\noindent

Using the averaging theorem (Sanders, 1985; Wiggins, 1996; Chicone et al., 1996a$\&$b 
and 1997a$\&$b) , we can now study the averaged dynamics of the system
(10) and (11) at resonance by introducing the averaging integration
\vskip  3pt
$$
{\bar {\cal Q}}\,=\,
{1\over {n{\cal T} _p}}\,{\int_0^{n{\cal T} _p}}\,{\cal Q}({{\cal I}_{_0}},
\Omega ({{\cal I}_{_0}})\,t+{\bar \Phi}\,,t)\,dt\,.
\eqno  (12)
$$
\vskip  6pt
\noindent
for any function ${\cal Q}({\cal I},\Theta,t)$. Also,
the averaging theorem requires $\dot {\cal D}$ and $\dot \Phi$ to have
the same power of perturbation parameter in the lowest order, that is 
${{\epsilon'}^2}\,=\,\epsilon\,.$ The partially averaged dynamics of the system (6) 
and (7) at resonance is now given by
$$
{\dot {\bar {\cal D}}}\,=\,{\epsilon^{1/2}}\,
{\bar {\cal P}}({{\cal I}_{_0}},{\bar \Phi},t)\,+\,
\epsilon\,{\bar {\cal D}}{\Bigl\langle}{{\partial {\cal P}} \over {\partial {\cal I}}}
({{\cal I}_{_0}},{\bar \Phi},t){\Bigr \rangle}\>,
\eqno  (13)
$$
\noindent
and
$$
{\dot{\bar \Phi}}\,=\,{\epsilon^{1/2}}\,{\bar {\cal D}}
{{\partial {\Omega}} \over {\partial {\cal I}}}({{\cal I}_{_0}})\,+\,
\epsilon\,\Bigl[{\bar {\cal N}}({{\cal I}_{_0}},{\bar \Phi},t)\,+\,
{1\over 2}\,{{\bar {\cal D}}^2}\,{\Bigl \langle}
{{{\partial^2}{\Omega}}\over {\partial {{\cal I}^2}}}({{\cal I}_{_0}})
{\Bigr \rangle} \Bigr]\>,
\eqno  (14)
$$
\noindent
where $\langle\>\rangle$ represent an averaged quantity.
Solutions to equations (13) and (14) are expected to represent the 
average dynamical behavior of the perturbation
system (6) and (7) over time intervals of order $\epsilon^{-1/2}$
(Sanders et al., 1985; Wiggins, 1996; Chicone et al., 1996a$\&$b and 1997a$\&$b).
\vskip  10pt
\noindent
{\centerline {\bigrmsixteen {4. $\>$ Second Order Partially Averaged System 
                                                               at Resonance}}}
\vskip  5pt

Mathematically speaking, equation (2) represents a dynamical system with
perturbation parameters $\varepsilon$ and $\delta$ (see Appendix A). In a parameter 
space, $\varepsilon$ and $\delta$ represent a plane where the full two-parameter
perturbation problem (2) must be studied. This, however, requires consideration
of all curves in this parameter space. However, in order to simplify the calculations 
and to avoid complexities, 
I consider a linear relation between $\varepsilon$ and $\delta$. 
That is, $\delta=\varepsilon \Delta$, with $\Delta$ fixed. This will simplify 
the analysis by reducing the two-parameter perturbation problem (2) to a
perturbation problem with only one parameter, $\varepsilon$.

At resonance, the orbital periods of Saturn and Jupiter become commensurate 
and therefore, the semimajor axis of Saturn's osculating ellipse becomes essentially 
constant. Following the formalism presented in
section 3, $L$ and $l$ can be written as
$$
L\,=\,{L_{_0}}\,+\,{\varepsilon^{1/2}}\,D\>\>\>,
\eqno  (15)
$$
\noindent
and
$$
l\,=\,{1\over {L_{_0}^3}}\,t\,+\,\varphi\>\>\>,
\eqno  (16)
$$
where $D$ represents the deviation of $L$ from its resonant value 
${L_{_0}}$ and $\varphi$ represents the deviation of $l$ 
from its resonant value ${L_{_0}^{-3}}\,t$.
Dynamics of the system near resonance is obtained by writing equation (2)
in terms of the new action-angle variables $(D,G,\varphi,g,)$.
The result is (paper I)
$$\!\!\!\!\!\!\!\!\!
\!\!\!\!\!\!\!\!\!\!\!\!\!\!\!\!\!\!\!\!\!\!\!\!\!\!\!\!\!\!\!\!
{\dot D}\>=\>-\,{\varepsilon^{1/2}}\>{F_{11}}\>-\>
\varepsilon D\>{F_{12}}\>+\>O({\varepsilon^{3/ 2}})\>\>\>,
\eqno  (17)
$$
$$
\!\!\!\!\!\!\!\!\!\!\!\!\!\!\!\!\!\!\!\!\!\!\!\!\!\!\!\!\!\!\!\!\!\!\!
\!\!\!\!\!\!\!\!\!\!\!\!\!\!\!\!\!\!\!\!\!\!\!\!\!\!\!\!\!\!\!\!\!\!\!\!\!
\!\!\!\!\!
{\dot G}\>=\>-\,\varepsilon\>{F_{22}}\>+\>O({\varepsilon^{3/ 2}})\>\>\>,
\eqno (18)
$$
$$
\!\!\!\!\!\!\!
{\dot \varphi}\>\>=\>-\>{\varepsilon^{1/2}}\>\Bigl({{3D}\over {L_{_0}^4}}\Bigr)
\>+\>\varepsilon\>\biggl({{6D^2}\over {L_{_0}^5}}\>+\>F_{32}\biggr)\>+\>
O({\varepsilon^{3/ 2}})\>\>\>,
\eqno (19)
$$
$$
\!\!\!\!\!\!\!\!\!\!\!\!\!\!\!\!\!\!\!\!\!\!\!\!\!\!\!\!\!\!\!\!\!\!\!
\!\!\!\!\!\!\!\!\!\!\!\!\!\!\!\!\!\!\!\!\!\!\!\!\!\!\!\!\!\!\!\!\!\!\!\!\!
\!\!\!\!\!\!\!\!\!\!
{\dot g}\>\>=\>\varepsilon\>F_{42}\>+\>O({\varepsilon^{3/ 2}})\>\>\>,
\eqno (20)
$$
\noindent
where
$$\!\!
{F_{11}}({L_{_0}},G,{L_{_0}^{-3}}\,t+\varphi,g) \,=\, 
{{\partial {{\cal H}_{ext}}} \over {\partial l}}\,({L_{_0}},G,{L_{_0}^{-3}}t+\varphi,g)
\,-\,{\Delta}\,{{\cal R}_L}({L_{_0}},G,{L_{_0}^{-3}}\,t+\varphi,g)\>\>\>,
\eqno  (21)
$$
$$\!\!
{F_{22}}({L_{_0}},G,{L_{_0}^{-3}}\,t+\varphi,g) \,=\, 
{{\partial {{\cal H}_{ext}}} \over {\partial g}}\,({L_{_0}},G,{L_{_0}^{-3}}\,t+\varphi,g)
\,-\,{\Delta}\,{{\cal R}_G}({L_{_0}},G,{L_{_0}^{-3}}\,t+\varphi,g)\>\>\>,
\eqno  (22)
$$
$$\!\!
{F_{32}}({L_{_0}},G,{L_{_0}^{-3}}\,t+\varphi,g) \,=\, 
{{\partial {{\cal H}_{ext}}} \over {\partial L}}\,({L_{_0}},G,{L_{_0}^{-3}}\,t+\varphi,g)
\,+\,{\Delta}\,{{\cal R}_l}({L_{_0}},G,{L_{_0}^{-3}}t+\varphi,g)\>\>\>,
\eqno  (23)
$$
$$\!\!
{F_{42}}({L_{_0}},G,{L_{_0}^{-3}}\,t+\varphi,g) \,=\, 
{{\partial {{\cal H}_{ext}}} \over {\partial G}}\,({L_{_0}},G,{L_{_0}^{-3}}\,t+\varphi,g)
\,+\,{\Delta}\,{{\cal R}_g}({L_{_0}},G,{L_{_0}^{-3}}\,t+\varphi,g)\>\>\>,
\eqno  (24)
$$
\noindent
and
$$\!\!\!\!\!\!\!\!\!\!\!\!\!\!\!\!\!\!\!\!\!\!\!\!
\!\!\!\!\!\!\!\!\!\!\!\!\!\!\!\!\!\!\!\!\!\!\!\!
\!\!\!\!\!\!\!\!\!\!\!\!\!\!\!\!\!\!\!\!\!\!\!\!
\!\!\!\!\!\!\!\!\!\!\!\!
{F_{12}}({L_{_0}},G,{L_{_0}^{-3}}\,t\,+\,\varphi\,,g)\,=
{{\partial {F_{11}}}\over {\partial L}}\,
({L_{_0}},G,{L_{_0}^{-3}}\,t\,+\,\varphi\,,g)\>\>\>.
\eqno  (25)
$$
\noindent

In equations (21)-(25), the effect of gravitational attraction of 
Jupiter appears in the external Hamiltonian ${\cal H}_{ext}$, where
$$
{{\cal H}_{ext}}\,=\,-\>{1\over {|{\vec r}\,-\,{{\hat r}_{_J}}|}}\>\>\>.
\eqno  (26)
$$
\noindent
Also the contribution of the frictional force of the interplanetary medium
is given by (Appendix C)
$$\!\!\!\!\!\!\!\!\!\!\!\!\!\!\!\!
{{\cal R}_L}\>=\,{a\over {\delta}}\>{(1-{e^2})^{-1/2}}\>
\bigl({{\cal R}_x}\,{\cal A}\,-\,{{\cal R}_y}\,{\cal B}\bigr)\>\>\>,
\eqno  (27)
$$
$$
{{\cal R}_G}\,=\,{1\over {\delta}}\,\Bigl[{{\cal R}_x}\,\sin (g\,+\,
{\hat v})\,-\,{{\cal R}_y}\,\cos\,(g\,+\,{\hat v})\Bigr]\>\>\>,
\eqno (28)
$$
$$\!\!\!\!\!\!\!\!\!\!\!\!\!\!\!\!\!\!\!\!\!\!\!\!\!\!\!
{{\cal R}_l}\>\>=\,-\,{r\over {e\delta}}\,{a^{-1/2}}\>\bigl({{\cal R}_x}
\,{\cal C}\,+\,{{\cal R}_y}\,{\cal W}\bigr)\>\>\>,
\eqno  (29)
$$
$$\!\!\!\!\!\!\!\!\!\!\!\!\!\!\!\!\!\!\!\!\!\!\!\!\!\!\!\!\!\!\!\!\!\!\!
\!\!\!\!\!\!\!\!\!\!\!\!\!\!
{{\cal R}_g}\>\>=\,{G\over {e\delta}}\>\bigl({{\cal R}_x}\,{\cal E}\,+\,
{{\cal R}_y}\,{\cal F}\bigr)\>\>\>,
\eqno  (30)
$$
\noindent
where
$$\!\!\!\!\!\!\!\!\!\!\!\!\!\!\!\!\!\!\!\!\!\!\!\!\!\!\!\!\!\!\!\!\!\!\!
\!\!\!\!\!\!\!\!\!\!\!\!\!\!\!\!\!\!\!\!\!\!\!\!\!\!\!\!\!\!\!\!\!\!\!
\!\!\!\!\!\!\!\!\!\!\!\!\!\!\!\!\!\!\!\!\!\!\!\!\!\!\!\!\!\!\!\!\!\!\!
\!\!\!\!\!\!\!\!\!\!\!\!\!\!\!\!\!\!\!\!\!\!\!\!\!\!\!\!\!\!\!
{\cal A}\,=\,\sin\,(g+{\hat v})\,+\,e\,\sin\,g\>\>\>,
\eqno  (31)
$$
$$\!\!\!\!\!\!\!\!\!\!\!\!\!\!\!\!\!\!\!\!\!\!\!\!\!\!\!\!\!\!\!\!\!\!\!
\!\!\!\!\!\!\!\!\!\!\!\!\!\!\!\!\!\!\!\!\!\!\!\!\!\!\!\!\!\!\!\!\!\!\!
\!\!\!\!\!\!\!\!\!\!\!\!\!\!\!\!\!\!\!\!\!\!\!\!\!\!\!\!\!\!\!\!\!\!\!
\!\!\!\!\!\!\!\!\!\!\!\!\!\!\!\!\!\!\!\!\!\!\!\!\!\!\!\!\!\!\!
{\cal B}\,=\,\cos\,(g+{\hat v})\,+\,e\,\cos\,g\>\>\>,
\eqno  (32)
$$
$$
{\cal C}\,=\,(-2e\,+\,\cos\,{\hat v}\,+\,e\,{\cos^2}{\hat v})\cos\,(g+{\hat v})
\,+\,(2\,+\,e\,\cos\,{\hat v})\sin\,(g+{\hat v})\,\sin \,{\hat v}\>\>\>,
\eqno  (33)
$$
$$
{\cal W}\,=\,(-2e\,+\,\cos\,{\hat v}\,+\,e\,{\cos^2}{\hat v})\sin\,(g+{\hat v})
\,-\,(2\,+\,e\,\cos\,{\hat v})\cos\,(g+{\hat v})\,\sin \,{\hat v}\>\>\>,
\eqno  (34)
$$
$$\!\!\!\!\!\!\!\!\!\!\!\!\!\!\!\!\!\!\!\!\!\!\!\!\!\!\!\!\!\!\!\!\!\!\!
\!\!\!\!\!\!\!\!\!\!\!\!\!\!\!\!\!\!\!\!\!\!\!\!\!\!\!\!\!\!\!\!\!\!\!
\!\!\!\!\!\!\!\!\!\!\!\!\!\!\!\!\!\!\!\!\!\!\!\!\!\!\!\!\!\!\!\!
{\cal E}\,=\,\cos \,g\,+\,\Bigl({{\sin {\hat v}}\over 
{1\,+\,e\,\cos\,{\hat v}}}\Bigr)\sin\,(g+{\hat v})\>\>\>,
\eqno  (35)
$$
$$\!\!\!\!\!\!\!\!\!\!\!\!\!\!\!\!\!\!\!\!\!\!\!\!\!\!\!\!\!\!\!\!\!\!\!
\!\!\!\!\!\!\!\!\!\!\!\!\!\!\!\!\!\!\!\!\!\!\!\!\!\!\!\!\!\!\!\!\!\!\!
\!\!\!\!\!\!\!\!\!\!\!\!\!\!\!\!\!\!\!\!\!\!\!\!\!\!\!\!\!\!\!\!
{\cal F}\,=\,\sin\,g\,-\,\Bigl({{\sin {\hat v}}\over 
{1\,+\,e\,\cos\,{\hat v}}}\Bigr)\cos\,(g+{\hat v})\>\>\>,
\eqno  (36)
$$
\noindent
and ${\cal R}_x$ and ${\cal R}_y$ are the Cartesian components of 
$\vec {\cal R}$. For the three-body system of Sun-Jupiter-Saturn in a 
(2:1) resonance, the dynamics of the second-order partially averaged system
at resonance is given by
$$\!\!\!\!\!\!\!\!\!\!\!\!\!\!\!\!\!\!\!\!\!\!\!\!\!\!
\!\!\!\!\!\!\!\!\!\!\!\!\!\!\!\!\!\!\!\!\!\!\!\!\!\!
\!\!\!\!\!\!\!\!\!\!\!\!\!\!\!\!\!\!\!\!\!\!\!\!\!
\!\!\!\!\!\!\!\!\!
{\dot {\bar D}}\,=\,-\,{\varepsilon^{1/2}}\,{{\bar F}_{11}}\,-\,
\varepsilon\,{\bar D}\>{{\bar F}_{12}}\>\>\>,
\eqno  (37)
$$
$$\!\!\!\!\!\!\!\!\!\!\!\!\!\!\!\!\!\!\!\!\!\!\!\!\!\!
\!\!\!\!\!\!\!\!\!\!\!\!\!\!\!\!\!\!\!\!\!\!\!\!\!\!
\!\!\!\!\!\!\!\!\!\!\!\!\!\!\!\!\!\!\!\!\!\!\!\!\!
\!\!\!\!\!\!\!\!\!\!\!\!\!\!\!\!\!\!\!\!\!\!\!\!\!\!\!\!\!\!
\!\!\!\!\!\!\!\!\!\!\!\!\!\!\!
{\dot {\bar G}}\,=\,-\,\varepsilon {{\bar F}_{22}}\>\>\>,
\eqno  (38)
$$
$$\!\!\!\!\!\!\!\!\!\!\!\!\!\!\!\!\!\!\!\!\!\!\!\!\!\!\!\!\!\!
\!\!\!\!\!\!\!\!\!\!\!\!\!\!\!\!\!\!\!\!\!\!\!\!\!
{\dot {\bar \varphi}}\,=\,-\,{\varepsilon ^{1/2}}\>
\Bigl({{3\bar D}\over {L_{_0}^4}}\Bigr)\,+\,\varepsilon\,
\Bigl({{6{{\bar D}^2}}\over {L_{_0}^5}}\,+\,
{{\bar F}_{32}}\Bigr)\>\>\>,
\eqno  (39)
$$
$$\!\!\!\!\!\!\!\!\!\!\!\!\!\!\!\!\!\!\!\!\!\!\!\!\!\!
\!\!\!\!\!\!\!\!\!\!\!\!\!\!\!\!\!\!\!\!\!\!\!\!\!\!
\!\!\!\!\!\!\!\!\!\!\!\!\!\!\!\!\!\!\!\!\!\!\!\!\!\!
\!\!\!\!\!\!\!\!\!\!\!\!\!\!\!\!\!\!\!\!
\!\!\!\!\!\!\!\!\!\!\!\!\!\!\!\!\!\!\!\!\!\!\!\!\!\!\!\!\!
{\dot {\bar g}}\,=\,\varepsilon\,{{\bar F}_{42}}\>\>\>.
\eqno  (40)
$$
\noindent
where
$$
{{\bar F}_{ij}}({\bar G}\,,\,{\bar g}\,,\,{\bar \varphi})\,=\,
{1\over {4\pi}}\>{\int_0^{4\pi}}\>{F_{ij}}\,
({\bar G}\,,\,{1\over {L_{_0}^3}}\,t\,+\,{\bar \varphi}\,,\,{\bar g}\,,\,t)\>dt\>\>\>.
\eqno (41)
$$

In order to study these equations, one needs to
calculate ${\bar F}_{ij}$. From equations (21)-(25), calculations of 
these terms involve the effect of gravitational attraction of Jupiter and
also the frictional effect of the interplanetary medium. Appendices B and C
contain these calculations, respectively.
From Appendix C and to the lowest order in eccentricity, the quantities 
${{\cal R}_L},{{\cal R}_G},{{\cal R}_l}$ and ${{\cal R}_g}$ are given by
\vskip  1pt
$$\!\!\!\!\!\!\!\!\!\!\!\!\!\!\!\!\!\!\!\!\!\!\!\!\!\!\!\!\!\!\!\!
\!\!\!\!\!\!\!\!\!\!\!\!\!\!
{{\cal R}_L}\,=\,-\,{1\over 2}\,{a^2}\,{e^2}\,\cos l\,
{\bigl(1-{3\over 4}\,{\cos ^2}\,l \bigr)^{1/2}}\>\>\>,
\eqno  (42)
$$
$$\!\!\!\!\!\!\!\!\!\!\!\!\!\!\!\!\!\!\!\!\!\!\!\!\!\!\!\!\!\!\!\!
\!\!\!\!\!\!\!\!\!\!\!\!\!\!\!\!\!
{{\cal R}_G}\,=\,-\,{1\over 2}\,a\,{e^2}\,\cos l\,
{\bigl(1-{3\over 4}\,{\cos ^2}\,l \bigr)^{1/2}}\>\>\>,
\eqno  (43)
$$
$$\!\!\!\!\!\!\!\!\!\!\!\!\!\!
{{\cal R}_l}\,=\,{1\over a}\,{e^2}\,
(137\,\sin l\,-\,23\,\sin 3l)\,
{\bigl(1-{3\over 4}\,{\cos ^2}\,l \bigr)^{1/2}}\>\>\>,
\eqno  (44)
$$
$$\!\!\!\!\!\!\!
{{\cal R}_g}\,=\,{1\over 8}\,G\,a\,{e^2}\,
(61\,\sin l\,+\,23\,\sin 3l)\,
{\bigl(1-{3\over 4}\,{\cos ^2}\,l \bigr)^{1/2}}\>\>\>.
\eqno  (45)
$$
\vskip  15pt
\noindent
Substituting for $l$ from equation (16) and using the averaging integration
(41), the averaged values of these quantities will be equal to zero.
 The non-vanishing contribution of friction appears as terms proportional 
to $e^3$. Since at resonance the numerical value of $e$ is only a few percent 
(paper I), the dynamical friction will not play a vital role in the dynamical 
evolution of the averaged system. Therefore, in the rest of these calculations, 
I neglect the contribution of dynamical friction at resonance
and will consider only the gravitational attraction of Jupiter as the main 
source of perturbation. Also, in order to simplify the
calculations, all quantities will be expanded only to the first order in
eccentricity.

Neglecting the frictional perturbation of the medium, the quantities ${\bar F}_{ij}$
will then be obtained only from the external Hamiltonian ${\cal H}_{ext}$.
From Appendix B these quantities are given by
\vskip  1pt
$$\!\!\!\!\!\!\!\!\!\!\!\!\!\!\!\!\!\!\!\!\!\!\!\!\!\!\!\!\!\!
\!\!\!\!\!\!\!\!\!\!\!\!\!\!\!\!\!\!\!\!\!\!\!\!\!\!\!\!
{{\bar F}_{11}}\,=\, {e\over {a_{_0}^2}}\,{\sigma_{11}}\,
\sin\,(2\varphi\,+\,g)\>\>\>,
\eqno  (46)
$$
$$\!\!\!\!\!\!\!\!\!\!\!\!\!\!\!\!\!\!\!\!\!\!\!\!\!\!\!\!\!\!
\!\!\!\!\!\!\!\!\!\!\!\!\!\!\!\!\!\!\!
{{\bar F}_{12}}\,=\,{1\over {e\,{a_{_0}^{5/2}}}}\,{\sigma_{11}}\,
\sin\,(2\varphi\,+\,g)\>\>\>,
\eqno  (47)
$$
$$\!\!\!\!\!\!\!\!\!\!\!\!\!\!\!
{{\bar F}_{22}}\,=\,{e\over {2{a_{_0}^2}}}\>
\Bigl[{\sigma_{22}^{(3/2)}}\,+\,3\,{\sigma_{22}^{(5/2)}}\Bigr]
\sin\,(2\varphi\,+\,g)\>\>\>,
\eqno  (48)
$$
$$\!\!\!\!\!\!\!\!\!\!\!\!\!\!\!\!\!\!\!\!\!\!\!\!\!\!\!\!\!\!\!\!\!\!\!\!
{{\bar F}_{32}}\,=\,-\,{1\over {2\,e\,{a_{_0}^{5/2}}}}\,{\sigma_{11}}\,
\cos\,(2\varphi\,+\,g)\>\>\>,
\eqno  (49)
$$
$$\!\!\!\!\!\!\!\!\!\!\!\!\!\!\!\!\!\!\!\!\!\!\!\!\!\!\!\!\!\!\!\!\!\!\!\!\!\!
\!\!\!\!\!\!\!\!
{{\bar F}_{42}}\,=\,{G\over {2e{a_{_0}^3}}}\,{\sigma_{11}}\,
\cos\,(2\varphi\,+\,g)\>\>\>,
\eqno  (50)
$$
\noindent
where
$$\!\!\!\!\!\!\!\!\!\!\!\!\!\!\!\!\!\!\!\!\!\!\!\!\!\!\!\!\!\!\!\!
\!\!\!\!\!\!\!\!\!\!\!\!\!\!\!\!\!\!\!\!\!\!
{\sigma_{11}}\>\>\>\>\>\>=\>{\sum_{h=0}^{\infty}}\>
{\Biggl[{{\Gamma({3\over 2}+h)}\over 
{{a_{_0}^h}\,{h!}\,{\Gamma({3\over 2})}}}\Biggr]^2}\>
\Biggl\{1\,+\,\Bigl({{2h+3}\over {h+1}}\Bigr)\>
\biggl[1\,-\,{3\over {4{a_{_0}^2}}}\>\Bigl({{2h+5}\over {h+2}}
\Bigr)\biggr]\Biggr\}\>\>\>,
\eqno  (51)
$$
$$\!\!\!\!\!\!\!\!\!\!\!\!\!\!\!\!\!\!\!\!\!\!\!\!\!\!\!\!\!\!\!\!
\!\!\!\!\!\!\!\!\!\!\!\!\!\!\!\!\!\!\!\!\!\!\!\!\!\!\!\!\!\!\!\!
\!\!\!\!\!\!\!\!\!\!
{\sigma_{22}^{(3/2)}}\,=\,{\sum_{h=0}^{\infty}}\>
{\Biggl[{{\Gamma({3\over 2}+h)}\over 
{{a_{_0}^h}\,{h!}\,{\Gamma({3\over 2})}}}\Biggr]^2}\>
\Biggl[1\,+\,{3\over {4{a_{_0}^2}}}\>\Bigl({{2h+5}\over {h+2}}\Bigr)\>
\Bigl({{2h+3}\over {h+1}}\Bigr)\Biggr]\>\>\>,
\eqno  (52)
$$
$$\!
{\sigma_{22}^{(5/2)}}\,=\,{\sum_{h=0}^{\infty}}\>
{\Biggl[{{\Gamma({5\over 2}+h)}\over 
{{a_{_0}^h}\,{h!}\,{\Gamma({5\over 2})}}}\Biggr]^2}\>
\Biggl\{1\,-\,{1\over {4{a_{_0}^2}}}\>\Bigl({{2h+5}\over {h+1}}\Bigr)\>
\biggl[3\,+\,\Bigl({{2h+7}\over {h+2}}\Bigr)\>
\Bigl(1\,-\,{3\over {4{a_{_0}^2}}}\>\>{{2h+9}\over {h+3}}\Bigr)\biggr]
\Biggr\}\>\>\>.
\eqno  (53)
$$
\noindent
Dynamical equations of the second-order averaged system at resonance can 
now be written as
$$\!\!\!\!\!\!\!\!\!\!\!\!\!\!\!\!\!\!\!\!\!\!\!\!\!\!\!\!\!\!
\!\!\!\!\!\!\!\!\!\!\!\!\!\!\!\!\!\!\!\!\!\!\!\!\!
{\dot D}\,=\,-\,{{e\,{\varepsilon^{1/2}}}\over {a_{_0}^2}}\>
\,{\sigma_{11}}\,\Bigl(1\,+\,{{\varepsilon^{1/2}}\over 
{{e^2}{a_{_0}^{1/2}}}}\,D\Bigr)\,\sin\,(2\varphi\,+\,g)\>\>\>,
\eqno  (54)
$$
$$\!\!\!\!\!\!\!\!\!\!\!\!\!\!\!\!\!\!\!\!\!\!\!\!\!\!\!\!\!\!
\!\!\!\!\!\!\!\!\!\!\!\!\!\!\!\!\!\!\!\!\!\!\!\!\!\!\!\!\!\!\!\!\!\!
{\dot G}\,=\,-\,{{e\,\varepsilon}\over {2\,{a_{_0}^2}}}\>
\Bigl[{\sigma_{22}^{(3/2)}}\,+\,3\,{\sigma_{22}^{(5/2)}}\Bigr]\>
\sin\,(2\varphi\,+\,g)\>\>\>,
\eqno  (55)
$$
$$\!\!\!\!\!\!\!\!\!\!\!\!\!\!\!\!\!\!\!\!\!\!\!\!\!\!\!\!\!\!\!\!\!\!\!\!
\!\!
{\dot \varphi}\,=\,-\,{{3\,\varepsilon^{1/2}}\over {a_{_0}^2}}\>D\,+\,
{{6\,\varepsilon}\over {a_{_0}^{5/2}}}\>
\Bigl[{D^2}\,-\,{1\over {12\,e}}\,\sigma_{11}\,\cos (2\varphi\,+\,g)
\Bigr]\>\>\>,
\eqno  (56)
$$
$$\!\!\!\!\!\!\!\!\!\!\!\!\!\!\!\!\!\!\!\!\!\!\!\!\!\!\!
\!\!\!\!\!\!\!\!\!\!\!\!\!\!\!\!\!\!\!\!\!\!\!\!\!\!\!\!\!\!
\!\!\!\!\!\!\!\!\!\!\!\!\!\!\!\!\!\!\!\!\!\!\!\!\!\!\!\!\!\!
\!\!\!\!\!\!\!\!\!\!\!\!\!\!\!\!\!\!
{\dot g}\,=\,{{G\,\varepsilon}\over {2e\,{a_{_0}^3}}}\,
{\sigma_{11}}\,\cos\,(2\varphi\,+\,g)\>\>\>.
\eqno  (57)
$$
\noindent
where the overbars have been dropped for the sake of simplicity.

Equations (54)-(57) allow us to analyze the averaged dynamics of the 
perturbation system (2) at resonance. Numerical integrations of these equations
indicate that the rate of change of orbital angular momentum of Saturn $(G)$
is almost zero (Figure 2) and its resonant value agrees with the value obtained
from the direct integrations of the main system (paper I).
\vskip  5pt
\hskip  100pt
$\!\!\!\!\!\!\!\!\!\!\!\!\!\!\!$
\epsfbox{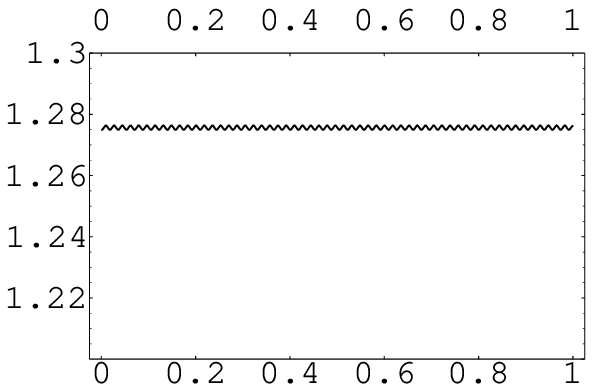}
\hfill
\vskip  1pt
\vbox{
\baselineskip=\normalbaselineskip
\smallrm 
\noindent
{\ti Figure 2.} {\smallrm {Graph of orbital angular momentum of Saturn versus
time at resonance. The initial conditions are given by 
$(D,G,\varphi,g)=(0,1.275,-5.4,0.882)$ and the timescale is 
${10^4}({T_{_J}}/{2\pi})$ years. }}}
\noindent

One can also use $G=L{(1-{e^2})^{1/2}}$
along with equation (15) to show that in the first order in eccentricity and 
second order in perturbation parameter $\varepsilon^{1/2}$, 
$$
{{de}\over {dt}}\,=\,{{G\,\varepsilon}\over {2{a_{_0}^3}}}\,
\Bigl[{\sigma_{22}^{(3/2)}}\,+\,3 {\sigma_{22}^{(5/2)}}\,-\,
2\,G\,{a_{_0}^{-1/2}}\,{\sigma_{11}}\Bigr]\,\sin (2\varphi+g)\>\>\>.
\eqno  (58)
$$
\noindent
Figure 3 shows that at resonance, the right hand side of equation (58)
is essentially zero which implies that eccentricity becomes almost constant
at resonance.
\vskip  50pt
\hskip 80pt
$\!\!\!\!\!\!\!\!\!\!\!\!\!\!\!$
\epsfbox{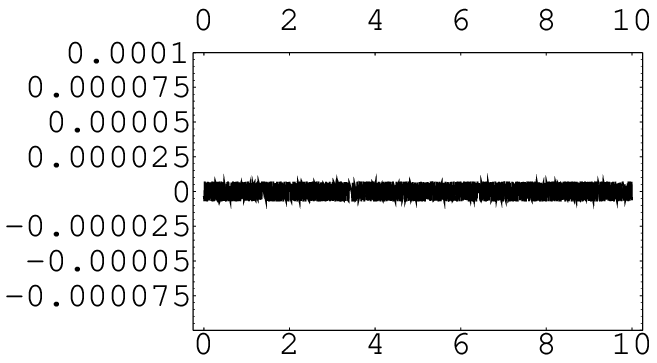}
\hfill
\vskip  1pt
\vbox{
\baselineskip=\normalbaselineskip
\smallrm 
\noindent
{\ti Figure 3.} {\smallrm {Graph of the rate of change of eccentricity
of Saturn's osculating ellipse versus time. This graph shows that at resonance 
$e$ is essentially constant. The initial conditions and the timescale are the 
same as Figure 2. }}}
\vskip  10pt
\noindent

Let us now turn our attention to evolution of action variable $D$ with time.
As mentioned 
in section 2, $D$ represents the deviation of $L$ from its Keplerian value.
One, therefore, expects $L$ and $D$ to have the same frequency.
In a comparison with
the result of the first order partially averaged system at resonance presented in 
paper I where the similarity between $D$ and $L$ was only pronounced for time
intervals of order $\varepsilon^{-1/2}$, the second-order partially averaged
equations reveal this similarity in general (Figure 4).
\vskip  5pt
{\centerline {\epsfbox{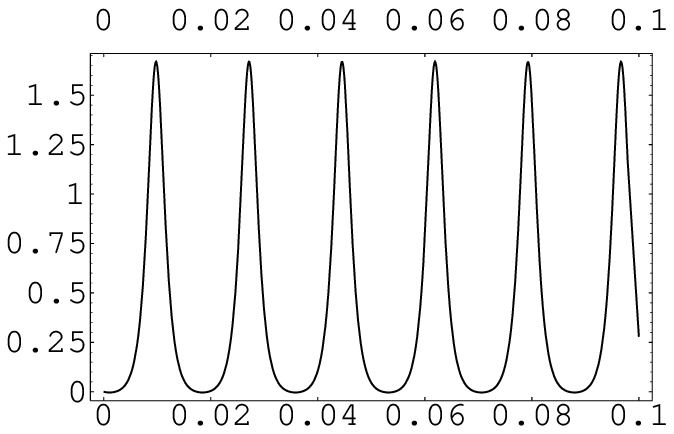}
\epsfbox{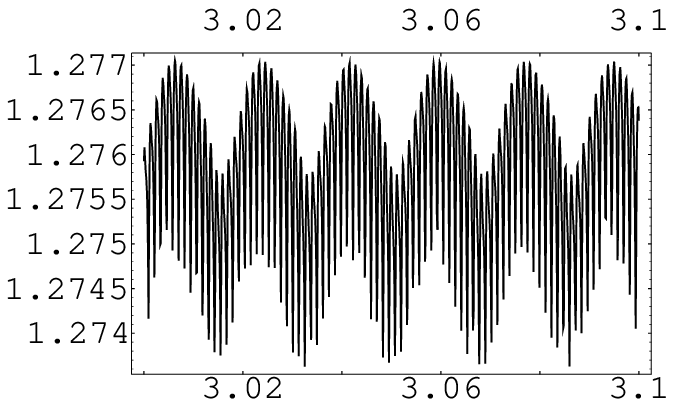}}}
\vskip  1pt
\vbox{
\baselineskip=\normalbaselineskip
\smallrm 
\noindent 
{\ti Figure 4.} {\smallrm {Graph of $D$ (left) and the action variable $L$ 
(right) versus time while the system is at resonance. The original system 
shows a periodic behavior (right) with a frequency which is in a
very good agreement with the second-order averaged system at resonance (left). 
The initial conditions for the graph of $D$ (left) are the same as Figure 2 and
for the graph of $L$ are the same as Figure 1. The timescale for both graphs is
${10^4}({T_{_J}}/{2\pi})$ years. }}}

It is necessary to mention that the graph of $L$ versus time shows two different 
frequencies (Figure 4). One is the frequency of
its envelope and the other is the Keplerian frequency of Saturn. From the
principle of averaging presented in section 3, it is the
frequency of the envelope that has to be similar to that of $D$. Figure 5
shows the frequency of the envelope and also the Keplerian frequency of
action variable $L$. The (2:1) commensurability is simply obtained by
calculating Saturn's Keplerian period from this figure and multiplying it
by the timescale ${10^4}({T_{_J}}/{2\pi})$. 
\hfill
\vskip  15pt
\hskip  40pt
\epsfbox{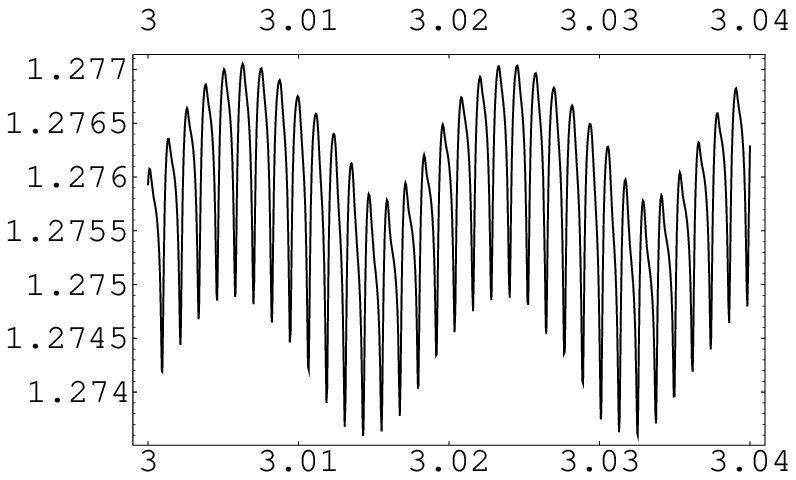}
\vskip  1pt
\vbox{
\baselineskip=\normalbaselineskip
\smallrm 
\noindent 
{\ti Figure 5.} {\smallrm {Graph of 
the action variable $L$ versus time while the system is at resonance. 
There are two frequencies associated with this quantity : A slow 
frequency related to the time variation of its envelope and a fast
frequency which is equal to the Keplerian frequency of Saturn at resonance.
The near (2:1) commensurability is simply obtained by dividing the time interval 
of this graph by the number of revolutions of Saturn during this interval and 
multiply the result by ${10^4}({T_{_J}}/{2\pi})$.}}}
\vskip  20pt
{\centerline {\bigrmsixteen { 5. $\>$ Summary}}}
\vskip  5pt

It has been shown by Haghighipour (1998) and in a more extensive way by
Melita and Woolfson (1996) that the three-body system of Sun-Jupiter-Saturn
will be captured in a near (2:1) resonance when one allows for the frictional
force of the interplanetary medium. In an attempt 
to study these results analytically, the method of partial averaging near 
a resonance was employed and the system was considered to be restricted,
planar and circular. The first-order partially averaged system 
confirmed $\>$ the $\>$ occurrence of $\>$ resonance 
capture for all initial relative 
positions of the two planets (Haghighipour, 1998); however, it was unable 
to fully illustrate the dynamical behavior of the system. The agreement
between evolutionary dynamics of the first-order partially averaged system
at resonance and the main system was observed only for time intervals of
order $({m_{_J}}/M)^{-1/2}$. In order to obtain a better agreement, dynamics
of the second-order partially averaged system at resonance has been studied. 
Numerical integrations of the averaged equations at this order have confirmed that 
the orbital angular momentum of Saturn and the eccentricity of its osculating ellipse
become essentially 
constant at resonance and demonstrated the close similarity between the dynamics 
of the averaged system at resonance and the main system over long time intervals.

The effect of dynamical friction has not appeared in this study. It has turned
out that the first non-vanishing terms of the averaged frictional force will be
proportional to $e^3$, which has been neglected in these calculations. 
It is important to mention that it is only through dynamical friction that
one can study the effect of the density of the interplanetary medium on
the dynamics of the system. However, calculations presented in this paper
indicate that considering dynamical friction as the only source of 
perturbation, the density of the medium will not play a
vital role in time variation of the orbital elements of Saturn at resonance. 
It is expected that accretion will provide the appropriate ground for the study 
of the effects of the density of the medium on resonance phenomena.

\vskip  7pt
\noindent
{\centerline{\bigrmsixteen {Acknowledgement}}}

I am deeply grateful to Dr.B.Mashhoon for critically reading the manuscript
and also for his valuable suggestions.
\vskip  1pt
\noindent
{\centerline {\bigrmsixteen {References}}}
\noindent
Binney, J. and Tremaine, S. : 1987, {\tibig {Galactic Dynamics}}, in Princeton
Series in  
\hfill
\vskip  1pt
\noindent
$\quad$Astrophysics, Princeton Univ. Press, Princeton, p.420.
\vskip  1pt
\noindent
Chicone, C. Mashhoon, B. and Retzloff, D.G. : 1996a, 
{\tibig {Ann.Inst.Henri Poincar\'e, Physique 
\vskip  1pt
\noindent
$\quad $Th\'erique,}} {\bf {64}}, 87.
\vskip  1pt
\noindent
Chicone, C. Mashhoon, B. and Retzloff, D.G. : 1996b,
{\tibig {J.Math.Phys.}}, {\bf {37}}, 3997.
\vskip  1pt
\noindent
Chicone, C. Mashhoon, B. and Retzloff, D.G. : 1997a, 
{\tibig {CQG}}, {\bf {14}}, 699.  
\vskip 1pt 
\noindent 
Chicone, C. Mashhoon, B. and Retzloff, D.G. : 1997b, 
{\tibig {CQG}}, {\bf {14}}, 1831.  
\vskip 1pt
\noindent 
Dodd, K.N. and McCrea, W.H. : 1952, {\tibig {MNRAS}}, {\bf {112}}, 205.  
\vskip  1pt
\noindent
Greenspan, B.D. and Holmes, P.J. : 1983, 
{\tibig {Nonlinear Dynamics and Turbulence}}, 
\vskip  1pt
\noindent
$\quad$ed. Barenblatt, G. Iooss, G. and Joseph, D.D.,
{\tibig {SIAM,  J.Math.Annal}}, {\bf {15}}, 69.
\vskip  1pt
\noindent
Guckenheimer, J. and Holmes, P.J. : 1983, 
{\tibig {Nonlinear Oscillations, Dynamical Systems
\vskip  1pt
\noindent
$\quad$and Bifurcation of Vector Fields,}} Springer-Verlag, New York.
\vskip 1pt
\noindent
Haghighipour, N. : 1998, to appear in {\tibig {MNRAS}},
(http://xxx.lanl.gov/abs/
\vskip  1pt
\noindent
$\quad$astro-ph/9812106)
\vskip  1pt
\noindent
Hagihara, Y. : 1972, {\tibig {Celestial Mechanics}}, Vol 2, MIT Press,  
Cambridge, p.267. 
\vskip  1pt
\noindent
Kovalevsky, J. : 1967, {\tibig {Introduction to Celestial Mechanics}},  
Springer-Verlag, New York.  
\vskip 1pt 
\noindent 
Melnikov, V.K. : 1963, {\tibig {Trans.Moscow.Math.Soc}}, {\bf {12}}, 1.
\vskip  1pt
\noindent
Sanders, J.A. and Verhulst, F. : 1985,
{\tibig {Averaging Methods in Nonlinear Dynamical Systems}}, 
\vskip  1pt
\noindent
$\quad$Springer-Verlag, New York.
\vskip  1pt
\noindent
Sternberg, S. : 1969, {\tibig {Celestial Mechanics}}, Vol 1, Benjamin, 
New York, p.110.
\vskip  1pt
\noindent
Wiggins, S. : 1996, {\tibig {Introduction to Applied Nonlinear Dynamical
Systems and Chaos}}, 
\hfill
\noindent
$\quad$Springer-Verlag, New York, p.143.  
\vskip  60pt
\noindent
{\centerline {\bigrmsixteen Appendix $\>$ A}}
\vskip  5pt
\noindent

Let us assume that the motions of Jupiter and Saturn are co-planar. 
In a plane-polar coordinate system with its origin on Sun, the equation
of motion of Saturn, i.e. equation (2) can be written as (paper I)
\vskip  1pt
$$\!\!\!\!\!\!\!\!\!\!\!\!\!\!\!\!\!\!\!\!\!\!\!\!\!\!\!\!\!\!\!\!\!\!\!\!
\!\!\!\!\!\!\!\!\!\!\!\!\!\!\!\!\!\!\!\!\!\!\!\!\!\!\!\!\!\!\!\!\!\!\!\!
\!\!\!\!\!\!\!\!\!\!\!\!\!\!\!\!\!\!\!\!\!\!\!\!\!\!\!\!\!\!\!\!\!\!\!\!
\!\!\!\!\!\!\!\!\!\!\!\!\!\!\!\!\!\!\!\!\!\!\!\!\!\!\!\!\!\!\!\!\!\!\!\!
\!\!\!\!\!\!\!\!\!\!\!\!\!\!\!\!\!\!\!\!\!\!\!\!\!\!\!\!\!\!\!\!\!\!\!\!
{P_r}\>=\> {\dot r}\>\>\>,
\eqno  (A1)
$$
$$\!\!\!\!\!\!\!\!\!\!\!\!\!\!\!\!\!\!\!\!\!\!\!\!\!\!\!\!\!\!\!\!\!\!\!\!
\!\!\!\!\!\!\!\!\!\!\!\!\!\!\!\!\!\!\!\!\!\!\!\!\!\!\!\!\!\!\!\!\!\!\!\!
\!\!\!\!\!\!\!\!\!\!\!\!\!\!\!\!\!\!\!\!\!\!\!\!\!\!\!\!\!\!\!\!\!\!\!\!
\!\!\!\!\!\!\!\!\!\!\!\!\!\!\!\!\!\!\!\!\!\!\!\!\!\!\!\!\!\!\!\!\!\!\!\!
\!\!\!\!\!\!\!\!\!\!\!\!\!\!\!\!\!\!\!\!\!\!\!\!\!\!\!\!\!\!\!
{P_\theta}\>=\> {r^2}\,{\dot \theta}\>\>\>,
\eqno  (A2)
$$
$$
{{\dot P}_r}\>=\>{{P_\theta^2}\over {r^3}}\>-\>{1\over{r^2}}\>-\>
{{\varepsilon}\over {|{\vec r}-{{\hat r}_{_J}}|^3}}\>
\big[r\,-\,\cos\,(\theta-{\theta_{_J}})\,\big]\>-\>
({{\cal R}_x}\cos\theta\>+\>{{\cal R}_y}\sin\theta)\>\>\>,
\eqno  (A3)
$$
\noindent
and
$$\!\!\!\!\!\!\!\!\!\!\!\!\!\!\!\!\!\!\!\!\!\!\!\!\!\!\!\!\!\!\!\!\!\!\!\!
\!\!\!\!\!\!\!\!\!\!\!\!\!\!\!\!\!
{{\dot P}_\theta}\,=\,-\,\varepsilon\,{{ r}\over {|{\vec r}-
{{\hat r}_{_J}}|^3}}\,
\sin(\theta-{\theta_{_J}})\,-\,
r(-\,{{\cal R}_x}\sin\theta\,+\,{{\cal R}_y}\cos\theta)\>\>\>,
\eqno  (A4)
$$
\vskip  1pt
\noindent
where
$$
{{\cal R}_x}\>=\> {A\over{{\cal V}_{rel}^3}}\>\ln \,
(1+B{r^2}{{\cal V}_{rel}^4})\>
\big[{\dot r}\cos\theta\,-\,
r({\dot \theta}-{\omega_m})\sin\theta\big]\>\>\>,
\eqno  (A5)
$$
$$
{{\cal R} _y}\>=\> {A\over{{\cal V}_{rel}^3}}\>\ln \,
(1+B{r^2}{{\cal V}_{rel}^4})\>
\big[{\dot r}\sin\theta\,+\,
r({\dot \theta}-{\omega_m})\cos\theta\big]\>\>\>.
\eqno  (A6)
$$
\vskip  5pt
\noindent

The Delaunay variables presented in section 4 are related to the plane-polar
coordinates by ${P_\theta^2}={G^2}=r(1+e\cos {\hat v})\,, 
{P_r}={{e\,\sin {\hat v}}/G}$ and $g=\theta-{\hat v}$. From these equations, 
the dynamical equations (A1)-(A4) can be written as
\vskip  1pt
$$\!\!\!\!\!\!\!\!\!\!\!\!\!\!\!\!\!\!\!\!\!\!\!\!\!\!\!\!\!\!\!\!\!\!\!
\!\!\!\!\!\!\!\!\!\!\!\!\!\!\!\!\!\!\!\!\!\!\!\!\!\!\!\!\!\!\!\!\!\!\!\!
{{dL}\over {dt}}\,=\,a\,(1-e^2)^{-{1/ 2}}\>\biggl[{F_r}\,e\,\sin{\hat v}\,+\,
{F_\theta}\,(1\,+\,e\,\cos{\hat v})\biggl]\>\>\>,
\eqno  (A7)
$$
$$\!\!\!\!\!\!\!\!\!\!\!\!\!\!\!\!\!\!\!\!\!\!\!\!\!\!\!\!\!\!\!\!\!\!\!
\!\!\!\!\!\!\!\!\!\!\!\!\!\!\!\!\!\!\!\!\!\!\!\!\!\!\!\!\!\!\!\!\!\!\!\!
\!\!\!\!\!\!\!\!\!\!\!\!\!\!\!\!\!\!\!\!\!\!\!\!\!\!\!\!\!\!\!\!\!\!\!\!
\!\!\!\!\!\!\!\!\!\!\!\!\!\!\!\!\!\!\!\!\!\!\!\!\!\!\!\!\!\!\!\!\!\!\!\!
\!\!\!\!\!\!\!\!\!\!\!\!\!\!\!\!\!\!\!\!\!\!\!\!\!\!\!\!\!\!\!\!\!\!\!\!
{{dG}\over {dt}}\,=\,r\,{F_\theta}\>\>\>,
\eqno  (A8)
$$
$$
{{dl}\over {dt}}\,=\,{\omega_{_K}}\,+\,{r\over e}\,{a^{-{1/ 2}}}\,
\Bigl[{F_r}\bigl(-2e\,+\,\cos {\hat v}\,+\,e\,{\cos^2} {\hat v}\bigr)\,-\,
{F_\theta}\,\bigl(2\,+\,e\,\cos {\hat v}\bigr)\,\sin {\hat v}\Bigr]\>\>\>,
\eqno (A9)
$$
\noindent
and
$$\!\!\!\!\!\!\!\!\!\!\!\!\!\!\!\!\!\!\!\!\!\!\!\!\!\!\!\!\!\!\!\!\!\!\!\!
{{dg}\over {dt}}\,=\,{1\over e}\,\Bigl[a(1-e^2)\Bigr]^{1/ 2}\,
\biggl[\,-{F_r}\cos {\hat v}\,+\,
{F_\theta}\Bigl(1\,+\,{1\over {1\,+\,e\,\cos{\hat v}}}\Bigl)
\sin{\hat v}\biggl]\>\>\>,
\eqno  (A10)
$$
\noindent
where 
$$
{F_r}\>=\>\varepsilon\>{{\cos(\theta-{\theta_{_J}})\,-\,r}\over {|{\vec r}-
{{\hat r}_{_J}}|^3}}\,-\,
({{\cal R}_x}\cos\theta\,+\,{{\cal R}_y}\sin\theta)\>\>\>,
\eqno  (A11)
$$
$$\!\!\!\!\!\!\!\!
{F_\theta}\>=\>-\,\varepsilon\,{{\sin(\theta-{\theta_{_J}})}\over {|{{\vec r}-
{{\hat r}_{_J}}|^3}}}\,+\,
({{\cal R}_x}\sin\theta\,-\,{{\cal R}_y}\cos\theta)\>\>\>,
\eqno  (A12)
$$
\vskip  4pt
\noindent
and the Keplerian frequency of the osculating ellipse is given by
${\omega_{_K}}\,=\,L^{-3}\>.$ The quantities $F_r$ and $F_\theta$ are associated 
with perturbations due to gravitational attraction of Jupiter and also 
dynamical friction of the interplanetary medium. 
\vskip  20pt
\noindent
{\centerline {\bigrmsixteen   Appendix $\>$ B }}
\vskip 5pt
\noindent

The external Hamiltonian ${\cal H}_{ext}$ can be written as
$$
{\cal H}_{ext}\,=\,-\,{\bigl[{r^2}\,-\,2r\cos (\theta\,-\,{\theta_{_J}})\,+\,
1\bigr]^{-{1/2}}}\>\>\>.
\eqno  (B1)
$$
\vskip  2pt
\noindent
Following the simplifications presented in paper I, quantities $r$ and
$\cos (\theta-{\theta_{_J}})$ are approximately given by
$$
r\,\simeq\,a\,(\,1\,-\,e\,\cos l)\>\>\>,
\eqno  (B2)
$$
\noindent
and
$$
\cos (\theta\,-\,{\theta_{_J}})\,\simeq\,\cos (l+g-{\theta_{_J}})\,+\,
e\,\Bigl[\cos (2l+g-{\theta_{_J}})\,-\,\cos (g-{\theta_{_J}})\,\Bigr]\>\>\>,
\eqno  (B3)
$$
\vskip  2pt
\noindent
where the $O({e^2})$ terms have been neglected. In this approximation, 
$$\eqalign {\!\!\!\!\!\!
{{\cal H}_{ext}}\simeq\,-\,&{\bigl[1+{a^2}-2a\cos (l+g-{\theta_{_J}})\bigr]^
{-{1/2}}}\cr
&\qquad\qquad\qquad
\biggl[1+{1\over 2}\>e\,a\>{{2\,a\cos l +\cos (2l+g-{\theta_{_J}})-
3\cos (g-{\theta_{_J}})}\over {1+{a^2}-2a\cos (l+g-{\theta_{_J}})}}\biggl]
\>\>\>.\cr}
\eqno  (B4)
$$
\vskip  2pt
\noindent
From equations (21)-(25), contribution of gravitational attraction of
Jupiter in dynamics of the second-order partially averaged system at resonance 
appears as partial
derivatives of ${\cal H}_{ext}$ with respect to $L,l,G$ and $g$. 
Substituting $a^2$ by $L$ and $e$ by ${(1-{G^2}/{L^2})^{1/2}}$, one will
obtain
$$
\eqalign {
{{\partial {{\cal H}_{ext}}}\over {\partial L}}\,\simeq\,
&{a^{1/2}}\>{\Bigl[1+{a^2}-2a\cos (l+g-{\theta_{_J}})\Bigr]^{-3/2}}\>
\Bigl[2a-{a\over e}\,\cos l\,+\,{3\over {2e}}\,
\cos (g-{\theta_{_J}})\cr
&\qquad\qquad\qquad\qquad\qquad\quad
-\,2\cos (l+g-{\theta_{_J}})-{1\over {2e}}\,
\cos(2l+g-{\theta_{_J}})\Bigr]\cr
&\!\!\!\!\!\!\!\!\!\!\!\!\!\!\!\!\!\!\!
+\,{3\over 2}\,e\,{a^{3/2}}\>
{\Bigl[1+{a^2}-2a\cos (l+g-{\theta_{_J}})\Bigr]^{-5/2}}\>
\Bigl[2(1+2{a^2})\,\cos\,l-8a\cos(g-{\theta_{_J}})\cr
&\qquad\qquad\qquad\qquad\qquad
-\,\cos (3l+2g-2{\theta_{_J}})\,+\,3\cos(l+2g-2{\theta_{_J}})\Bigr]\>\>\>,\cr}
\eqno  (B5)
$$
\vskip  1pt
$$
\eqalign {
{{\partial {{\cal H}_{ext}}}\over {\partial l}}\,\simeq\,
&a\,{\Bigl[1+{a^2}-2a\cos (l+g-{\theta_{_J}})\Bigr]^{-3/2}}
\>\Bigl[ea\sin l+\sin (l+g-{\theta_{_J}}) \cr
&\qquad\qquad\qquad\qquad\qquad\qquad\qquad\qquad\qquad\qquad
+ e\sin (2l+g-{\theta_{_J}})\Bigr]\cr
&+\,{3\over 2}\,e{a^2}\>
{\Bigl[1+{a^2}-2a\cos (l+g-{\theta_{_J}})\Bigr]^{-5/2}}
\,\sin (l+g-{\theta_{_J}})\cr
&\qquad\qquad\qquad\qquad\qquad
\Bigl[2a\cos l+\cos (2l+g-{\theta_{_J}})
-3\cos (g-{\theta_{_J}})\Bigr]\>\>\>,\cr}
\eqno  (B6)
$$
\vskip  1pt
$$
\eqalign {
{{\partial {{\cal H}_{ext}}}\over {\partial G}}\,\simeq\,{G\over {2\,e}}\>
&{\Bigl[1+{a^2}-2a\cos (l+g-{\theta_{_J}})\Bigr]^{-3/2}}\cr
&\qquad\qquad\qquad
\Bigl[2\,a\,\cos l + \cos (2l+g-{\theta_{_J}})-
3\cos (g-{\theta_{_J}})\Bigr]\>\>\>,\cr}
\eqno  (B7)
$$
and
$$
\eqalign {
{{\partial {{\cal H}_{ext}}}\over {\partial g}}\,\simeq\,&a\,
{\Bigl[1+{a^2}-2a\cos (l+g-{\theta_{_J}})\Bigr]^{-3/2}}\cr
&\qquad
\biggl\{\sin (l+g-{\theta_{_J}})\,+\,
{1\over 2}\,e\,\Big[\sin\,(2l+g-{\theta_{_J}})\,-\,
3\sin(g-{\theta_{_J}})\Bigr]\biggr\}\cr
&\!\!\!\!\!\!\!\!\!
-\,{3\over 4}\,e\,{a^2}\>
{\Bigl[1+{a^2}-2a\cos (l+g-{\theta_{_J}})\Bigr]^{-5/2}}\cr
&\qquad
\Bigl[4\,\sin l\,-\,2a\,\sin(g-{\theta_{_J}})+
3\sin (l+2g-2{\theta_{_J}})\cr
&\qquad\qquad\qquad\qquad
-\,2a\sin (2l+g-{\theta_{_J}})\,-\,\sin (3l+2g-2{\theta_{_J}})
\Bigr]\>\>\>.\cr}
\eqno  (B8)
$$
\vskip  3pt

Quantities above can be simplified noting that at resonance ${a_{_0}}\simeq 1.625$
(paper I). Therefore one can use the relation
$$
{\bigl(\,1\,-\,2\xi \cos \alpha\,+\,{\xi^2}\bigr)^{-\lambda}}\,=\,
{\sum_{q=0}^\infty}\>{C_q^\lambda}\,(\cos \alpha)\>{\xi^q}\qquad ,
\qquad |\xi|\,<\,1\>\>\>,
\eqno  (B9)
$$
\vskip  2pt
\noindent
with $\xi={a_{_0}^{-1}}$ and $\alpha=(l+g-{\theta_{_J}})$, to write equations 
(B5)-(B8) in a simple form. In this relation, ${C_q^\lambda}(\cos \alpha)$ are 
Gegenbauer polynomials which are given by
\vskip  1pt
$$
{C_q^\lambda}\,(\cos \alpha)\,=\,{\sum_{h=0}^q}\>
{{\Gamma (\lambda+h)\>\Gamma (\lambda+q-h)}\over
{h!\>(q-h)!\>{[\Gamma (\lambda)]^2}}}\> \cos\big[(q-2h)\,\alpha\,\big]\>\>\>.
\eqno  (B10)
$$
\vskip  10pt
\noindent

It is important to note that integration (41) applies certain restrictions
to the admissible values of $q$ and $h$. Due to  the harmonic nature of the
terms produced by equation (B10), the only terms that survive the integration
(41) are the ones proportional to $\cos (2l+g-{\theta_{_J}})$ and
$\sin (2l+g-{\theta_{_J}})$. On the other hand, expansion of quantities
(B5)-(B8) using Gegenbauer polynomials, 
will reslut in producing terms proportional to
$\cos [(q-2h)(l+g-{\theta_{_J}})]$ and $\sin [(q-2h)(l+g-{\theta_{_J}})]$. 
After simplifying equations (B5) to (B8), one realizes that the only terms 
that can have non-zero contributions are the ones with angular arguments
equal to $l\,,\,(g-{\theta_{_J}})\,,\,(l+2g-2{\theta_{_J}})\,,\,
(2l+g-{\theta_{_J}})\,$ and
$\,(3l+2g-2{\theta_{_J}})$ which require $q$ to be equal to $(2h \pm 1)\,,\,
(2h \pm 2)\,,\, (2h\pm 3)\,,(2h)\, $ and $\,(2h \pm 1)$, respectively. 
Denoting the contributing 
part of the external Hamiltonian by ${\cal H}_{ext}^\ast$,  the non-vanishing 
terms in equations (B5)-(B8) can be written as
\vskip  10pt
\noindent
$$\!\!\!\!\!\!\!\!\!\!\!\!\!\!\!\!\!\!\!\!\!\!
{{\partial {{\cal H}_{ext}^\ast}}\over {\partial L}}\,\simeq\,
-\,{1\over {2\,e\,{a_{_0}^{5/2}}}}\,{\sigma_{11}}\>
\cos (2l+g-{\theta_{_J}})\>\>\>,
\eqno  (B11)
$$
$$\!\!\!\!\!\!\!\!\!\!\!\!\!\!\!\!\!\!\!\!\!\!\!\!\!\!\!\!\!\!\!\!\!\!\!\!
\!\!\!\!
{{\partial {{\cal H}_{ext}^\ast}}\over {\partial l}}\,\simeq\,{e\over {a_{_0}^2}}\>
{\sigma_{11}}\>\sin (2l+g-{\theta_{_J}})\>\>\>,
\eqno  (B12)
$$
$$
{{\partial {{\cal H}_{ext}^\ast}}\over {\partial g}}\,\simeq\,
{e\over 2{{a_{_0}^2}}}\>\bigr[{\sigma_{22}^{(3/2)}}\,+\,
3\,{\sigma_{22}^{(5/2)}}\bigl]\>\sin (2l+g-{\theta_{_J}})\>\>\>,
\eqno  (B13)
$$
\vskip  2pt
\noindent
and
$$\!\!\!\!\!\!\!\!\!\!\!\!\!\!\!\!\!\!\!\!\!\!\!\!\!\!\!\!\!\!
{{\partial {{\cal H}_{ext}^\ast}}\over {\partial G}}\,\simeq\,
{G\over {2\,e\,{a_{_0}^3}}}\,{\sigma_{11}}\,
\cos (2l+g-{\theta_{_J}})\>\>\>.
\eqno  (B14)
$$
\vskip  10pt
\noindent
where ${\sigma_{11}}\>,{\sigma_{22}^{(3/2)}}$ and $\sigma_{22}^{(5/2)}$ 
are given by
equations (51)-(53). Substituting equations (B11)-(B14) in equations
(21)-(25) and integrating the results using integration (41), 
${{\bar F}_{11}},{{\bar F}_{12}},{{\bar F}_{22}},{{\bar F}_{32}}$ and
${{\bar F}_{42}}$ can be written as in equation (46)-(50).
\vskip  20pt
\noindent
{\centerline {\bigrmsixteen  Appendix $\>$ C} }

I have shown in paper I that equations of motion in terms of the Delaunay 
variables can also be written as
\vskip  1pt
$$
{{dL}\over {dt}}\>=\>-\,\varepsilon\,{{\partial {{\cal H}_{ext}}}\over 
{\partial l}}\>+\>\varepsilon \, \Delta \,{{\cal R}_L}\>\>\>,
\eqno  (C1)
$$
$$
{{dG}\over {dt}}\>=\>-\,\varepsilon\,
{{\partial {{\cal H}_{ext}}}\over {\partial g}}
\>+\>\varepsilon\,\Delta\,{{\cal R}_G}\>\>\>,
\eqno  (C2)
$$
$$\qquad
{{dl}\over {dt}}\>\>\>=\,{1\over{L^3}}\,+\,\varepsilon \,
{{\partial {{\cal H}_{ext}}}\over {\partial L}}
\>+\>\varepsilon \, \Delta\,{{\cal R}_l}\>\>\>\>\>\>,
\eqno  (C3)
$$
\noindent
and
$$
\!\!\!
{{dg}\over {dt}}\>=\,\varepsilon
{{\partial {{\cal H}_{ext}}}\over {\partial G}}
\>+\>\varepsilon\,\Delta\,{{\cal R}_g}\>\>\>\>\>\>,
\eqno  (C4)
$$
\vskip  6pt
\noindent
Comparing equations (C1)-(C4) with equations (A7)-(A10), 
${{\cal R}_L},{{\cal R}_G},{{\cal R}_l}$ and ${\cal R}_g$ can be written as in 
equations (27)-(30). After substituting for ${\cal R}_x$ and ${\cal R}_y$
from equations (A5) and (A6) in these equations,
the effect of the frictional force of the
interplanetary medium on dynamics of the averaged system appears as
$$\!
{{\cal R}_L}=-{a\over {B{{\cal V}_{rel}^3}}}\,{(1-{e^2})^{-1/2}}\,
\Bigl[a\,(1-{e^2})\,({\dot \theta}-{\omega_m})+e{\dot r} \sin {\hat v}\,\Bigr]
\ln (1+B {r^2}{{\cal V}_{rel}^4})\>\>\>,
\eqno  (C5)
$$
$$\!\!\!\!\!\!\!\!\!\!\!\!\!\!\!\!\!\!\!\!\!\!\!\!
\!\!\!\!\!\!\!\!\!\!\!\!\!\!\!\!\!\!\!\!\!\!\!\!\!\!\!\!
\!\!\!\!\!\!\!\!\!\!\!\!\!\!\!\!\!\!\!\!\!\!\!\!\!\!\!\!
\!\!\!\!\!\!\!\!\!\!\!\!\!\!\!\!\!\!\!\!\!\!\!\!\!\!\!\!
\!\!\!\!\!\!\!\!\!\!\!\!\!\!\!\!\!\!
{{\cal R}_G}=-{r\over {B{{\cal V}_{rel}^3}}}\,({\dot \theta}-{\omega_m})
\,\ln(1+B {r^2}{{\cal V}_{rel}^4})\>\>\>,
\eqno  (C6)
$$
$$\!
{{\cal R}_l}=-{{r{a^{-1/2}}}\over {eB{{\cal V}_{rel}^3}}}\>
\Bigl[\,{\dot r}\,(-2e+\cos{\hat v}+e{\cos^2}{\hat v})-
r\,({\dot \theta}-{\omega_m})\,(2+e\cos {\hat v})\,\sin {\hat v}\,\Bigr]
\ln(1+B{r^2}{{\cal V}_{rel}^4})\,,
\eqno  (C7)
$$
$$\!
{{\cal R}_g}={G\over {eB{{\cal V}_{rel}^3}}}
\Bigl[\,{\dot r}\cos {\hat v}-r({\dot \theta}-{\omega_m})
\Bigl(1+{1\over {1+e\cos {\hat v}}}\Bigr)\sin {\hat v}\,\Bigr]
\ln(1+B{r^2}{{\cal V}_{rel}^4})\>\>\>.
\eqno  (C8)
$$
\vskip  10pt
\noindent
I have shown in paper I that at resonance, the numerical value of 
the logarithmic term $\ln\,(1\,+\,B\,{r^2}\,{{\cal V}_{rel}^4})$ becomes
so small that one can use the approximation 
$\>\ln (1+\zeta)\simeq \zeta\>,\>\zeta<<1$ to
simplify equations (C5)-(C8). Using this approximation and also 
substituting for ${\omega_m},{\dot \theta},{\dot r}$ and ${\cal V}_{rel}$ by
(paper I)
\vskip 1pt
$$\!\!\!\!\!\!\!\!\!\!\!
{\omega_m}\,\simeq\,{a^{-3/2}}\,\Bigl(\,1\,+\,{3\over 2}\,e\,\cos l\,
\Bigr)\>\>\>,
\eqno  (C9)
$$
$$\!\!\!\!\!\!\!\!\!\!\!\!
{\dot \theta}\>\>\>\simeq\,{a^{-3/2}}\>\Bigl(\,1\,+\,2\,e\cos l\,
\Bigr)\>\>\>,
\eqno  (C10)
$$
$$\!\!\!\!\!\!\!\!\!\!\!\!\!\!\!\!\!\!\!\!\!\!\!\!\!\!\!\!\!\!\!\!\!\!\!
{\dot r}\>\>\simeq\,e\,{a^{-1/2}}\>{\sin l}\>\>\>.
\eqno  (C11)
$$
\noindent
and
$$
{{\cal V}_{rel}}\,\simeq\,{a^{-1/2}}\,e\,\bigl(1\,-\,{3\over 4}\,{\cos ^2} l
\Bigr)^{1/2}\>\>\>,
\eqno  (C12)
$$
\vskip  2pt
\noindent
respectively, ${{\cal R}_L},{{\cal R}_G},{{\cal R}_l}$ and ${\cal R}_g$
can be written as in equations (42)-(45).

\end